\documentclass[aps,prs,nofootinbib,twocolumn]{revtex4-1}
\usepackage{amsmath,amssymb,graphics,graphicx}

\newcommand{\be}{\begin{equation}} \newcommand{\ee}{\end{equation}}

\begin{document}
\title
{\Large \bf Nonzero $\theta_{13}$ signals nonmaximal atmospheric neutrino mixing}

\author{David A. Eby\footnote{daeby@physics.unc.edu} 
and  Paul H. Frampton\footnote{frampton@physics.unc.edu}}

\affiliation{Department of Physics and Astronomy, University of North Carolina, 
Chapel Hill, NC 27599-3255, USA}

\date{\today}

\begin{abstract}
From recent groundbreaking experiments,
it is now known that the Pontecorvo-Maki-Nakagawa-Sakata
mixing differs significantly from the 
tribimaximal model in which $\theta_{13}= 0$
and $\theta_{23} = \pi/4$. 
Flavor symmetry can require that the departures
from these two equations are linearly related.
$T^{'}$ and $A_4$, which successfully accommodated the pre-T2K
Pontecorvo-Maki-Nakagawa-Sakata matrix, predict that
$38.07^{\circ} \leq \theta_{23} \leq 39.52^{\circ}$
at 95\% C.L.. The best fit values, combining the model predictions
with T2K, MINOS, Double Chooz, Daya Bay, and RENO data, 
are $\theta_{23}=38.7^{\circ}$ and $\theta_{13}=8.9^{\circ}$.
\end{abstract}


\maketitle

Of the parameters in the standard model 
of particle theory, we will focus on the mixing
matrices for down-type quarks and for neutrinos, named
respectively for Cabibbo, Kobayashi, and Maskawa (CKM)~\cite{C,KM} 
and for Pontecorvo, Maki, Nakagawa, and Sakata (PMNS)~\cite{P,MNS}.
Without losing generality, we choose a basis in
which the flavor and mass eigenstates coincide for
the three up-type quarks and all three charged leptons.

This investigation will consider one of three mixing angles of CKM quark mixing 
($\Theta_{12}$) and two of
the three
mixing angles of PMNS neutrino mixing 
($\theta_{13}$ and $\theta_{23}$), ignoring for the moment the {\it CP}-violating 
phases in both cases.

We recall the values of the angles $\theta_{13}$
and $\theta_{23}$ listed in the 2010 Review of Particle Physics\footnote{The reader is directed to the references summarized in RPP.}
\cite{RPP2010} since these two are, we suggest, both changed
by the T2K measurement \cite{T2Ktalk, T2K, Dufour:2011zz, Hartz:2012np, Frank:2011zz, Izmaylov:2011np}. 
The values then were
\begin{equation}
36.8^{\circ} \lesssim \theta_{23} \leq 45.0^{\circ} ,  ~~~~~~~~~~~~~~  
0.0^{\circ} \leq \theta_{13} \lesssim 11.4^{\circ}
\label{2313}
\end{equation}
consistent with vanishing $\theta_{13}$ and maximal $\theta_{23}$.

The other angles
are not considered to be variables in this analysis, although the superior experimental accuracy of the CKM Gell-Mann-L\'{e}vy quark mixing angle~\cite{GL},
\begin{equation}
\Theta_{12} = (13.03 \pm 0.06)^{\circ},
\label{GMC}
\end{equation}
played an important role in our investigation of flavor symmetry.

To accommodate the new data, we invoke 
broken binary tetrahedral ($T^{'}$) flavor symmetry as a promising approach 
to explaining the mixing angles~\cite{FKtprime,FK2001,Feruglio:2007uu,Frampton:2007et,
Frampton:2008bz,CM,FKR,EFM1,EFM2,FHKM}.

This flavor symmetry was 
first used in Ref.~\cite{FKtprime} solely as a symmetry
for quarks, because
neutrinos were still believed to be massless. After neutrino masses and
mixings were discovered~\cite{SuperK}, the mixing matrix for neutrinos 
was measured and found to be very different from
the CKM mixing matrix for quarks. A number of theories arose 
~\cite{Frampton:1993wu,Aranda:2000tm,He:2003rm,Babu:2005se,Lee:2006pr} to 
explain this. Eventually, a useful approximation to the empirical PMNS mixing was 
determined to be the tribimaximal (TBM) matrix~\cite{HPS},

\begin{equation}
U_{TBM} = \left( \begin{array}{ccc}
\sqrt{2/3} & \sqrt{1/3} & 0 \\
- \sqrt{1/6} & \sqrt{1/3} & - 1/\sqrt{2} \\ 
- \sqrt{1/6} & \sqrt{1/3} & 1/\sqrt{2} 
\end{array} \right).
\label{TBM}
\end{equation}

Flavor symmetry based on the tetrahedral group, $A_4 = T$,
was introduced by Ref.~\cite{Ma} to underpin
TBM neutrino mixing. Further investigation revealed that this model could not be extended to
quarks because a viable CKM matrix could not be obtained~\cite{Altarelli}.
$A_4$
is not a subgroup of its double cover~\cite{FKR}, $T^{'}$,
nevertheless from the viewpoint of kronecker products
used in model building~\cite{FK2001}, $A_4$ behaves
{\it as if} it were a subgroup. This explains why 
the larger group can act as a successful
flavor symmetry for both quarks and leptons.

\begin{figure*}
\includegraphics[scale=0.47]{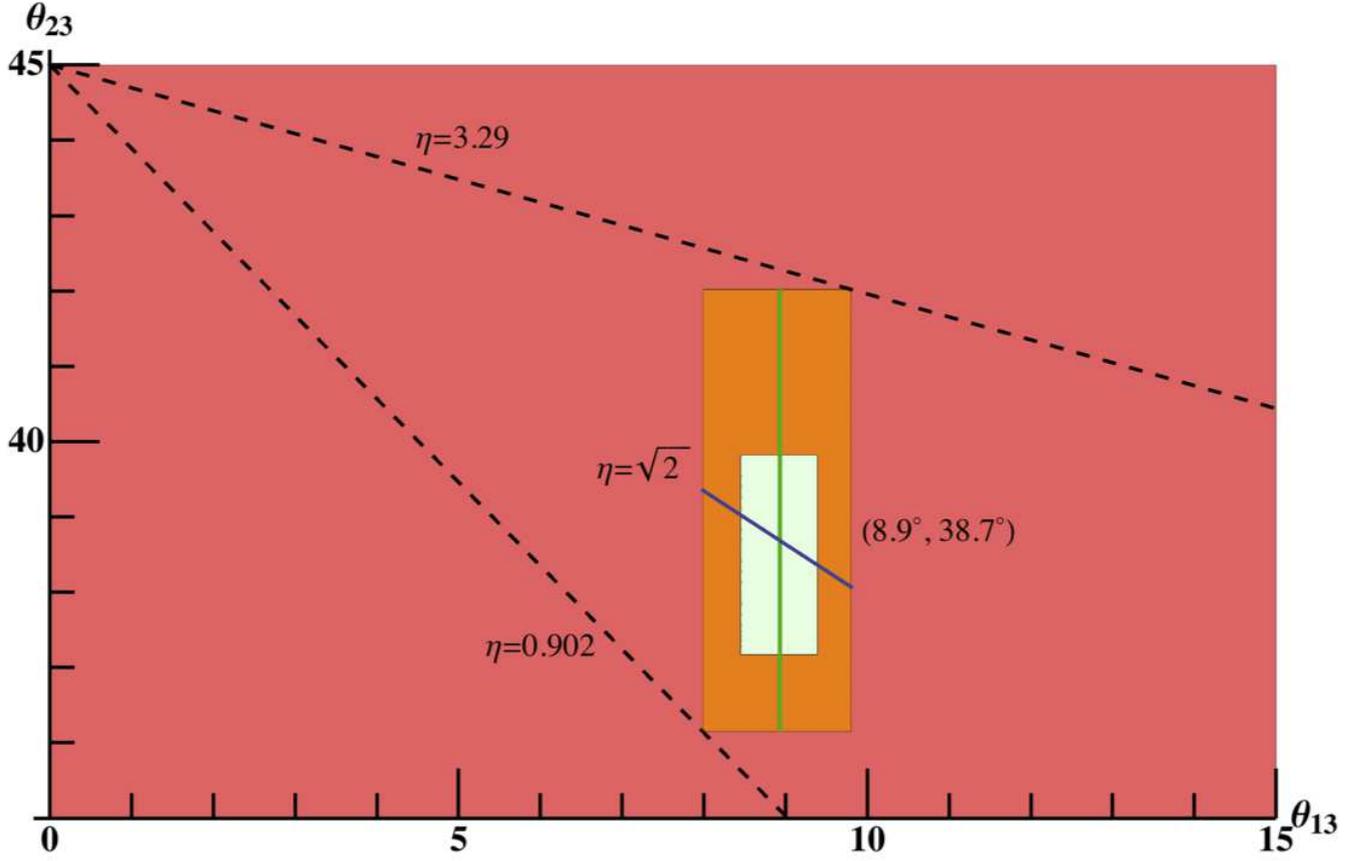}%Two Column Scale [scale=0.45]
\caption{
The global analysis of Ref.~\cite{Fogli:2012ua}, incorporating SBL, LBL, solar, and atmospheric neutrino observations, excludes the red-shaded region at $2\sigma$. The same assessment excludes the orange-shaded region at $1\sigma$. The best fit value for $\theta_{13}$ is indicated by the vertical green line at $\theta_{13}=8.9^{\circ}$. Extreme values of the linear correlation coefficient, $\eta$, are indicated by dashed lines at $\eta=0.902$ and $\eta=3.29$, while our predicted correlation of $\eta=\sqrt{2}$ is indicated by the solid dark blue line. The intersection of our correlation prediction and the $\theta_{13}$ best fit occurs at $\theta_{13} = 8.9^{\circ}$ and $\theta_{23} = 38.7^{\circ}$, a close match to the current experimental best fit of $\theta_{23}=38.4^{\circ}$. 
}
\label{figure1}
\end{figure*}

We shall consider only the projection on the
two-dimensional $\theta_{23}$ - $\theta_{13}$ plane of the three-dimensional
$\theta_{12}$~-~$\theta_{23}$~-~$\theta_{13}$ space. At leading
order, requiring $\sin{\alpha} \sim \alpha$\footnote{This is a $ < 1\%$  approximation for 
$\theta_{13}$ and $(\frac{\pi}{4} - \theta_{23})$ since both angles are less 
than $\alpha = 12^{\circ} = 0.2094$ radians with $\sin{\alpha}~=~0.2079$.} 
for $\theta_{13}$ and $(\frac{\pi}{4} - \theta_{23})$,
the calculation of the perturbation of this projection
from the TBM matrix in Eq.(\ref{TBM})
is independent of the solar neutrino mixing angle $\theta_{12}$.
The relevant perturbation away from
Eq.(\ref{TBM}) was explicitly calculated 
in Ref.~\cite{Frampton:2008bz,EFM1}.

Before T2K, the neutrino mixing angles 
were all empirically consistent with the TBM values. However,
as the experimental accuracy has now  improved in recent
data from T2K~\cite{T2K, Dufour:2011zz, Hartz:2012np, Frank:2011zz, Izmaylov:2011np, T2Ktalk}, 
MINOS~\cite{MINOStalk, Adamson:2012rm, MINOS, Holin:2012np, Habig:2011zz, Orchanian:2011qq, Evans:2010zza}, Double Chooz~\cite{DCtalk, DC, Palomares:2011zz, Abe:2011fz, Palomares:2009wz}, Daya Bay~\cite{Daya, DBtalk}, and RENO~\cite{RENO, RENOtalk}, this situation has
changed dramatically, as discussed in the global fits of 
Refs.~\cite{Tortola:2012te, Fogli:2012ua, GonzalezGarcia:2012sz}; of these we shall use Fogli {\it et al}.~\cite{Fogli:2012ua}. These five remarkable experiments have provided us with a rich new perspective on mixing angles.
From flavor symmetry, it is then possible to predict quantitatively
how departures from the TBM values,
\begin{equation}
\theta_{12} = \tan^{-1} \left( \frac{1}{\sqrt{2}} \right),
~~ \theta_{23} = \left( \pi/4 \right), ~~~ \theta_{13} =0,
\label{TBM2}
\end{equation}
are related. The 
model allows one to address this question by relating
the perturbations around TBM,
\begin{equation}
\theta_{ij} = (\theta_{ij})_{TBM} + \epsilon_k,
\label{PerturbedTBM}
\end{equation}
(where $\epsilon_3$ corresponds to $\theta_{12}$, and so on)
to the analogous perturbations around the minimal model's
prediction for the CKM Gell-Mann-L\'{e}vy quark mixing angle,
\begin{equation}
\tan{2(\Theta_{12})}  = \left( \frac{\sqrt{2}}{3} \right).
\label{CabibboTprime}
\end{equation}

The data from KamLAND, LBL accelerators (like T2K and MINOS), solar experiments, SBL accelerators (such as Double Chooz, Daya Bay, and RENO), and Super-Kamiokande, as combined in Ref.~\cite{Fogli:2012ua}
indicate (accounting for CP violation)
\begin{equation}
\sin^2{\theta_{13}}=0.0241_{~-0.0048}^{~+0.0049}~~~~~
{\rm with}~95\%~{\rm C.L.}
\label{t2k}
\end{equation}
for a normal neutrino mass hierarchy, as favored by $T^{'}$.

Because Eq. (\ref{CabibboTprime}) yields a value of $\Theta_{12} = 12.62^{\circ}$, 
which while close is significantly below the experimental
value, Eq. (\ref{GMC}), it is possible to perturb
to the empirical $\Theta_{12}$ and to track the deviations
in the PMNS mixing matrix to the linear relationship,\footnote{$A_4$ is also capable of producing Eq.(\ref{eta}) with $\eta = \sqrt{2}$, though we give preference in this paper to $T^{'}$ for its capacity to explain CKM mixing.}
\begin{equation}
\theta_{13} = \eta \left( \frac{\pi}{4} - \theta_{23} \right),
\label{eta}
\end{equation} 
with the sharp 
prediction\footnote{It is notable that Eq.(\ref{eta}) with
$\eta \simeq \sqrt{2}$ appears {\it en passant}
in Ref.~\cite{HS};
see also Ref.~\cite{FukiYasue} which implies that $\eta \sim 2$. 
Another, model-independent correlation was developed in Ref.~\cite{Ge:2011qn}, 
including the three PMNS mixing angles and the {\it CP}-violating phase.} 
that $\eta = \sqrt{2}$. Thus,
\begin{equation}
\theta_{13} = \sqrt{2} \left( \frac{\pi}{4} - \theta_{23} \right),
\label{eta2}
\end{equation} 
This prediction is derived in further detail in Ref.~\cite{EFM1}.

Considering the result, Eq. (\ref{eta}),
it requires that, if $\eta$ is finite as expected, any departure from 
$\theta_{13} = 0$ signals that $\theta_{23} < \pi/4$.
As shown in Fig. (\ref{figure1}), the recent
experimental data, combined with theory,
suggest that ($\theta_{13}$, $\theta_{23}$) are respectively
closer to ($8.9^{\circ}$, $38.7^{\circ}$)
than to ($0.0^{\circ}$, $45.0^{\circ}$). 
Before T2K, $\eta$ was unconstrained, $0 \leq \eta < \infty$.
With the current global fit data, we find $0.902 \leq \eta \leq 
3.29$.

This is in sharp contrast to the previously widespread 
acceptance of a maximal $\theta_{23} = \pi/4$,
which fitted so well with vanishing $\theta_{13} = 0$
in the TBM context.

As the measurement of $\theta_{13}$ sharpens experimentally, so
will the prediction for $\theta_{23}$ from Eq.~(\ref{eta2}), and 
measurement of the atmospheric
neutrino mixing's departure from maximality will provide
an interesting test of the binary tetrahedral flavor symmetry.

Several years ago Super-Kamiokande showed $\theta_{23} > 36.8^{\circ}$~\cite{SK},
and current analysis places it at $\theta_{23} \simeq 40.7^{\circ}$~\cite{SKtalk}. Once 
combined in a global fit of $3 \nu$ oscillation, Ref.~\cite{Fogli:2012ua} states the best 
fit of $\theta_{23}=38.4^{\circ}$, tantalizingly close to our central value of 
$\theta_{23} = 38.7^{\circ}$.

This suggests to us that
the $T^{'}$ flavor symmetry, introduced in Ref.~\cite{FKtprime}, 
should now be taken much more seriously. As errors in $\theta_{13}$ 
and $\theta_{23}$ diminish even further, it will be interesting to see how 
the prediction of Eq.~(\ref{eta2}) by $T^{'}$ perseveres, as it would 
inspire further investigation into other mixing angles for quarks and 
leptons. This, in turn, may show that $T^{'}$, first mentioned in physics 
as an example of an $SU(2)$ subgroup~\cite{Yang}, is actually a 
useful approximate symmetry in the physical application of quark and 
lepton flavors.

\bigskip

\noindent
This work was supported by DOE Grant No. 
DE-FG02-05ER41418 and DOE-GAANN Award No.
P200A090135.

\end{document}